\documentclass[prl,twocolumn,aps,superscriptaddress,showpacs,floatfix]{revtex4}

\usepackage{amsmath}
\usepackage{amssymb}
\usepackage{amsbsy}
\usepackage{graphicx}
\usepackage{color}
\usepackage{epstopdf}
\usepackage{mathrsfs}

\begin{document}
	
	\title{Exact Mobility Edges and Topological Phase Transition in Two-Dimensional non-Hermitian Quasicrystals}
	
	\author{Zhihao Xu}
	\affiliation{Institute of Theoretical Physics and State Key Laboratory of Quantum Optics and Quantum Optics Devices, Shanxi University, Taiyuan 030006, China}
	\affiliation{Beijing National Laboratory for Condensed Matter Physics, Institute of Physics, Chinese Academy of Sciences, Beijing 100190, China}
	\affiliation{Collaborative Innovation Center of Extreme Optics, Shanxi University, Taiyuan 030006, China}
	
	\author{Xu Xia}
	\email{xiaxu14@mails.ucas.ac.cn}
	\affiliation{Chern Institute of Mathematics and LPMC, Nankai University, Tianjin 300071, China}
	
	\author{Shu Chen}
	\email{schen@iphy.ac.cn}
	\affiliation{Beijing National Laboratory for Condensed Matter Physics, Institute of Physics, Chinese Academy of Sciences, Beijing 100190, China}
	\affiliation{School of Physical Sciences, University of Chinese Academy of Sciences, Beijing, 100049, China}
	\affiliation{Yangtze River Delta Physics Research Center, Liyang, Jiangsu 213300, China}
	
	\begin{abstract}
		The emergence of the mobility edge (ME) has been recognized as an important characteristic of Anderson localization. The difficulty in understanding the physics of the MEs in three-dimensional (3D) systems from a microscopic image encourages the development of models in lower-dimensional systems that have exact MEs. While most of the previous studies are concerned with one-dimensional (1D) quasiperiodic systems, the analytic results that allow for an accurate understanding of two-dimensional (2D) cases are rare. In this work, we disclose an exactly solvable 2D quasicrystal model with parity-time ($\mathcal{PT}$) symmetry displaying exact MEs. In the thermodynamic limit, we unveil that the extended-localized transition point, observed at the $\mathcal{PT}$ symmetry breaking point, is topologically characterized by a hidden winding number defined in the dual space. The coupling waveguide platform can be used to realize the 2D non-Hermitian quasicrystal model, and the excitation dynamics can be used to detect the localization features.
	
\end{abstract}

	\maketitle
	
	\section{Introduction}\label{section1}
		
		Anderson localization \cite{Anderson}, illustrated by the absence of wave propagation in a disordered medium, is an essential phenomenon in condensed-matter physics, which has been observed in various wave systems \cite{Cutler,Ramakrishnan,Dalichaouch,Wiersma,Storzer,Topolancik,Schwartz,Strybulevych,Chabe,Sperling,Lopez,Manai,Tianping,Hutchinson}. The localization properties of a disordered system are determined by its dimensionality \cite{Abrahams}. The one-parameter scaling theory \cite{Abrahams} predicts that all states in one- and two-dimensional (1D and 2D) systems are localized at infinitesimal random disorder regimes. For a 3D case, beyond a critical disorder amplitude, a mobility edge (ME), which marks critical energy separating extended states from localized ones, may lead to some fundamental physics \cite{Evers}, such as the metal-insulator transition and the strong thermoelectric response \cite{Whitney,Chiaracane}. Nevertheless, understanding the ME in 3D systems from a microscopic image is difficult, which promotes exploring the models with MEs in lower-dimension, especially with exact MEs.
		
		Replacing the random disorder with a quasiperiodic one, which hosts localized and extended states even in the low-dimension regime, is one way to reproduce the similar localization transition of higher-dimensional disordered systems in lower dimensions. These low-dimensional quasicrystals (QCs) display metal-insulator transitions and even MEs predicted in 1D QCs \cite{AA,Sarma1988,Boers,Biddle2009,Biddle2010,Lellouch,XLi2017,HYao2019,Ganeshan,ZhihaoXu2020,YLiu2020,QBZeng,Longhi2021,XuXia,YanxiaoLiu2021,LiuPRB2021,LingZhiTang,LiuTong,LonghiPRB2021,LonghiPRB20211,TongLiu2020,LonghiPRB2019,LonghiPRL2019,HJiang,XDeng,Yucheng,ZhihaoXuPRA,YuCheng2021}. A paradigmatic example of a 1D QC is provided by the Aubry-Andr\'{e} (AA) model, in which the metal-insulator transition can be derived from a self-duality argument without MEs \cite{AA}. However, by introducing an energy-dependent self-duality, one can obtain the generalized AA models with exact MEs, such as the 1D QCs with a long-range hopping term \cite{Biddle2010} or a unique form of the on-site incommensurate potential \cite{Ganeshan}. In particular, with the development of the studying of the non-Hermitian systems in experimental and theoretical fields \cite{Bender,Bender1,Hatano1,Hatano2,Hatano3,SLongi1,HuitaoShen,Shunyu1,Shunyu2,Zongping,Lee1,Lee2,Harari,Parto,HengyunZhou,Ruter,LiangFeng,Regensburger,Joglekar,XiangZhan,LeiXiao,Zeuner,YongXu,Okuma,LinhuLi,Kunst,Takata,MPan,Nakagawa,Hamazaki,Yamamoto,Ashida,Kawabata,Kawabata2,Zirnstein,Okuma1,Longwen1,Longwen2}, one has devoted to the interplay of non-Hermiticity and disorder, which brings a new perspective of the localization features \cite{Hatano1,Hatano2,Hatano3,Zongping}. The 1D non-Hermitian QCs with exact MEs have been investigated \cite{YLiu2020,Longhi2021,XuXia,YanxiaoLiu2021,YuCheng2021}.
		
		The majority of previous ME research has focused on 1D cases; however, little is known about that in the 2D QCs, though some numerical evidence shows the existence of MEs in 2D systems \cite{BHuang,Pupillo,Schneider,Asatryan}. This work aims to disclose the exact MEs in 2D QCs, described by the 2D parity-time ($\mathcal{PT}$) symmetric QCs with two incommensurate modulation frequencies along the $x$- and $y$-directions. For the 2D non-Hermitian QCs, we derive the exact expression of the MEs and the localization transition points analytically. It is also shown that one of the localization transition points, detected at the $\mathcal{PT}$ symmetry breaking point in the thermodynamic limit, is topologically characterized by a hidden winding number. The excitation dynamics are used to investigate the localization properties of 2D non-Hermitian QCs realized in an optical waveguide platform.
		
		\section{Model and Hamiltonian}\label{sec:2}
		We consider a 2D non-Hermitian QC model, which can be described by the Hamiltonian
		\begin{align}\label{eq1}
			\hat{H}(\lambda) = & -t\sum_{m,n}(\hat{c}^{\dagger}_{m,n}\hat{c}_{m+1,n}+\hat{c}^{\dagger}_{m,n}\hat{c}_{m,n+1}+H.c.) \notag \\ & +\sum_{m,n}\lambda_{mn}\hat{c}^{\dagger}_{m,n}\hat{c}_{m,n}.
		\end{align}
		Here, $\hat{c}_{m,n}$ is the annihilation operator at the $(m,n)$-th site in 2D lattices with $m$ and $n$ corresponding to the position of the lattice along the $x$- and $y$-directions, respectively, and $m,n=1,2,\dots,L$ with $L\times L$ being the total number of the lattices. $t$ is the hopping amplitude between the nearest neighbors, which is set $t=1$ as unit energy. The complex on-site potential
		\begin{equation}\label{eq2}
			\lambda_{mn} = \lambda e^{i 2\pi(\alpha_x m + \alpha_y n)},
		\end{equation}
		where $\lambda$ is the modulation amplitude ($\lambda\ge 0$), $\alpha_{x}$ and $\alpha_y$ are irrational for a 2D QC, which control the modulation frequencies along the $x$- and $y$-directions, respectively. In this work, we fix our discussion on $\alpha_{x}/\alpha_{y}$ being an irrational number. Note that one has $\lambda_{-m-n}=\lambda_{mn}^*$, and $H(\lambda)$ is $\mathcal{PT}$ symmetric. Suppose that the eigenstate of a single-particle in the 2D lattices is given by $|\psi_{m,n}\rangle = \sum_{m,n}\psi_{m,n} \hat{c}^{\dagger}_{m,n}|0\rangle$, the eigenenergies of the system can be obtained from the eigenvalue equation:
		\begin{equation}\label{eq3}
			\psi_{m,n+1}+\psi_{m,n-1}+\psi_{m+1,n}+\psi_{m-1,n}=(\lambda_{mn}-\varepsilon)\psi_{m,n},
		\end{equation}
		where $\psi_{m,n}$ is the amplitude of the single-particle wave function at the $(m,n)$-th site and $\varepsilon$ is the single-particle eigenvalue. In the following calculation, we focus on the localization properties and the topological phase transition of the 2D $\mathcal{PT}$ symmetric quasiperiodic system with $\alpha_{x}/\alpha_{y}$ being an irrational number.
		
		\section{Localization transitions and exact MEs}\label{sec:3}
		We perform the Sarnak method \cite{Sarnak} to analytically study the localization features of the 2D non-Hermitian QC model, of which the eigenvalue equation is described by Eq. (\ref{eq3}). Let us introduce the Fourier transformation
		\begin{equation}\label{eq4}
			f(\theta_x,\theta_y)=\frac{1}{L}\sum_{m,n} \psi_{m,n}e^{i (m \theta_x + n \theta_y)},
		\end{equation}
		with $\theta_{\beta}=2\pi \alpha_{\beta} l_{\beta}, \beta=\{x,y\}$, and $l_{\beta}$ being the site index in the dual space along the $\beta$-direction. In the thermodynamic limit, the dual equation is given as
		\begin{equation}\label{eq5}
			\left[\varepsilon+2\cos{\theta_x}+2\cos{\theta_y}\right]f(\theta_x,\theta_y)=\lambda f(\theta_x+\bar{\omega}_x,\theta_y+\bar{\omega}_y),
		\end{equation}
		with $\bar{\omega}_{\beta}=2\pi\alpha_{\beta}$. We define a characteristic function as follows:
		\begin{equation}\label{eq6}
			G(\varepsilon)=\frac{1}{(2\pi)^2}\int_0^{2\pi}d\theta_x\int_{0}^{2\pi}d\theta_y\ln{|g(\varepsilon,\theta_x,\theta_y)|},
		\end{equation}
		with
		\begin{equation}\label{eq7}
			g(\varepsilon,\theta_x,\theta_y)=\varepsilon+2\cos{\theta_x}+2\cos{\theta_y}.
		\end{equation}
		The Sarnak method states the following: for a given $\varepsilon$, (i) $G(\varepsilon)<\ln{\lambda}$, $\varepsilon$ is not the solution of the eigenvalue equation; (ii) $G(\varepsilon)=\ln{\lambda}$, the spectrum is dense with localized states and $\varepsilon$ being a complex value; (iii) $\{G(\varepsilon)>\ln{\lambda}\} \cap U_{\varepsilon}$, it corresponds to a dense set of $\varepsilon$ with extended states, where $U_{\varepsilon}=[-4,4]$ is the range of spectrum such that $g(\varepsilon,\theta_x,\theta_y)=0$ for some $\theta_x,\theta_y$ and $\varepsilon$ is a real value. For the eigenvalue in the region $U_{\varepsilon}$, $G(\varepsilon)\in [0,1.1662]$. Hence, there are two metal--insulator transition points at $\lambda_{c1}=e^{0}=1$ and $\lambda_{c2}=e^{1.1662}\approx 3.2098$. When $\lambda<\lambda_{c1}$, all the eigenstates are extended, and the corresponding eigenvalues are within $[-4,4]$. In the regime $\lambda>\lambda_{c2}$, all the eigenstates are localized, and all the eigenvalues satisfy $G(\varepsilon)=\ln{\lambda}$. For the case of $\lambda\in[\lambda_{c1},\lambda_{c2}]$, the system displays the MEs $\pm \varepsilon_{c}$, which are real numbers that satisfy
		\begin{equation}\label{eq8}
			G(\pm\varepsilon_{c})=\ln{\lambda}.
		\end{equation}
		The eigenstates with the real part of eigenvalues $\mathrm{Re}(\varepsilon)\in [-\varepsilon_c,\varepsilon_c]$ are localized, while all the eigenstates are extended with $\varepsilon \in [-4,-\varepsilon_c]\cup[\varepsilon_c,4]$ in $\lambda\in[\lambda_{c1},\lambda_{c2}]$. Thus, our results exactly exhibit the localization features of the 2D non-Hermitian quasiperiodic system.
		
		\begin{figure}[tbp]
			\begin{center}
				\includegraphics[width=.5 \textwidth] {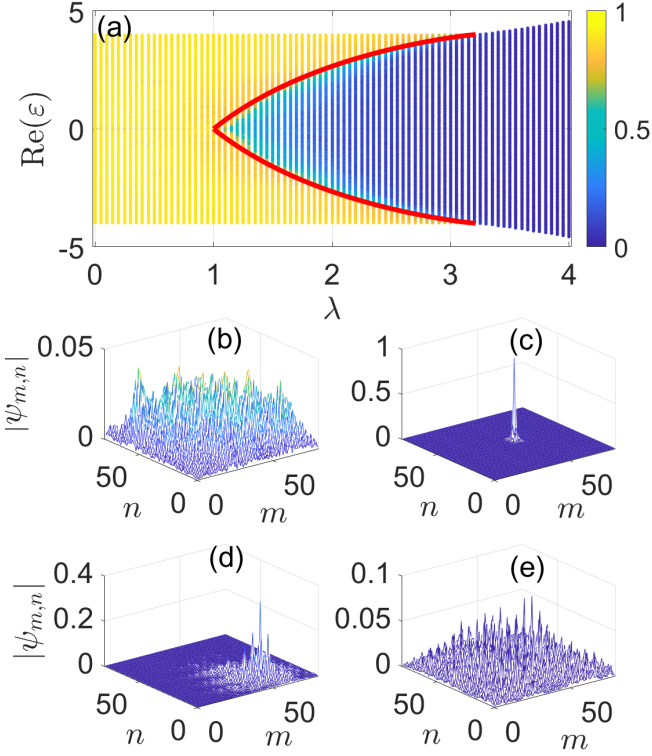}
			\end{center}
			\caption{(Color online) (a) Fractal dimension $\eta$ of different eigenstates as a function of the corresponding real parts of the eigenvalues $\mathrm{Re}(\varepsilon)$ and the quasiperiodic potential amplitude $\lambda$. The red solid lines represent the MEs given by $G(\pm \varepsilon_c)=\ln{\lambda}$ with $\varepsilon_c$ being a real value. (b)-(e) Profiles of different eigenstates correspond to (b) $\lambda=0.8$ and $\mathrm{\varepsilon}\approx 0$, (c) $\lambda=3.3$ and $\mathrm{\varepsilon}\approx 0$, (d) $\lambda=2.2$ and $\mathrm{\varepsilon}\approx 2.7$, and (e) $\lambda=2.2$ and $\mathrm{\varepsilon}\approx 3.0$, respectively. Here, we fix $\alpha_{x}=(\sqrt{5}-1)/2$, $\alpha_y=\sqrt{3}$, $L=80$ and OBCs.}\label{Fig1}
		\end{figure}
		
		To numerically verify our analytical results, we can calculate the fractal dimension for an arbitrary $j$-th normalized eigenstates $\psi^{(j)}$ with $\sum_{m,n}|\psi^{(j)}_{m,n}|^2=1$, which is given by $\eta=-\lim_{L\to\infty} \ln{(\mathrm{IPR})}/\ln{(L^2)}$ for a 2D system with the system size $L\times L$, with the inverse participation ratio (IPR) being $\mathrm{IPR}_{j}=\sum_{m,n}|\psi^{(j)}_{m,n}|^4$. It is known that $\eta\to 1$ for extended states and $\eta \to 0$ for localized states. Figure \ref{Fig1}(a) shows the fractal dimension $\eta$ of different eigenstates as a function of the corresponding real parts of eigenvalues $\mathrm{Re}(\varepsilon)$ and the modulation amplitude $\lambda$ under open boundary conditions (OBCs). The red solid lines in Fig. \ref{Fig1}(a) represent the MEs $\pm \varepsilon_c$ solution of Eq. (\ref{eq8}) in $\lambda\in [\lambda_{c1},\lambda_{c2}]$. As expected from our analytical results, when the eigenvalues across the MEs, $\eta$ approximately turns from zero to one. The localized properties can be further confirmed by the profile of the eigenstate $|\psi_{m,n}|$ with different $\lambda$ and $\mathrm{Re}(\varepsilon)$ which is shown in Figs. \ref{Fig1}(b)-(e). When $\lambda<\lambda_{c1}$ [Fig. \ref{Fig1}(b) for $\lambda=0.8$], the eigenstate with $\mathrm{Re}(\varepsilon)\approx 0$ is extended. In $\lambda>\lambda_{c2}$ regime, taking $\lambda=3.3$ and  $\mathrm{Re}(\varepsilon)\approx 0$ as an example shown in Fig. \ref{Fig1}(c), the profile of the eigenstate exhibits a localized feature. The profiles of the eigenstates with $\mathrm{Re}(\varepsilon)\approx 2.7$ and $3.0$ are localized and extended shown in Figs. \ref{Fig1}(d) and \ref{Fig1}(e), which, respectively, correspond to eigenvalues below and above the ME $\varepsilon_{c}\approx 2.98$ for $\lambda=2.2$. To strengthen the calculation of localization properties, we further study the scaling behaviors of different localization regimes shown in Appendix A1.
		
		\begin{figure}[tbp]
			\begin{center}
				\includegraphics[width=.5 \textwidth] {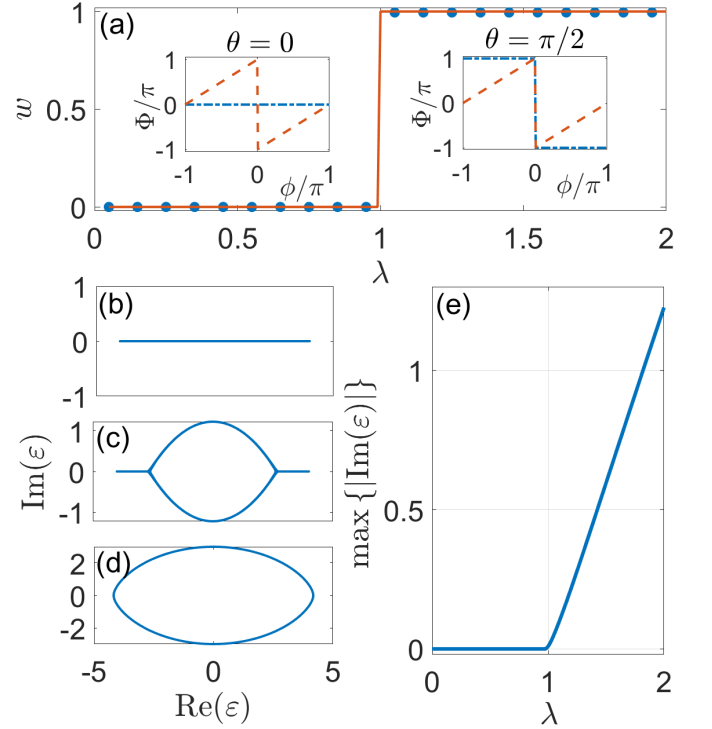}
			\end{center}
			\caption{(Color online) (a) Winding number $w$ versus $\lambda$ with $\varepsilon_B=0$ for the numerical calculation (circles) and the analytical results (solid line), respectively, and the data are averaged $10000$ different $\theta$ realizations. The insets show $\Phi$ as a function of $\phi$ for $\theta=0$ and $\theta=\pi/2$, respectively. The dash-dot lines denote $\lambda=0.95$ and the dash lines represent $\lambda=1.05$. (b)-(d) Energy spectrum with rolling $\theta$ from $0$ to $2\pi$, and $\lambda=0.8$, $2.0$ and $3.5$, respectively. (e) Behavior of the maximal value of $|\mathrm{Im}(\varepsilon)|$ versus $\lambda$ for varying $\theta$ from $-\pi$ to $\pi$. Here, we numerically calculate the effective system described by Eq. $\eqref{eq9a}$ with $\alpha_{x}=(\sqrt{5}-1)/2$, $\alpha_y=\sqrt{3}$, $L=1000$ and under PBCs.}\label{Fig2}
		\end{figure}

		\section{Topological phase transition and $\mathcal{PT}$-symmetry breaking}
		The topological features of our 2D $\mathcal{PT}$ symmetric QC model can be unveiled in its dual space. By using the Fourier transformation Eq. (\ref{eq4}), we obtain the dual equation Eq. (\ref{eq5}) of the original 2D QCs. In the thermodynamic limit, one notices that the dual equation along the diagonal direction can be decoupled to a series of 1D non-Hermitian chains for a given $l=l_x-l_y$ with the eigenvalue equations as follows:
		\begin{align}
			\lambda f_{l_x+1}-2\left[\cos{(2\pi \alpha_x l_x)}+\cos{(2\pi \alpha_{y}l_x-\theta)}\right]f_{l_x} &=\varepsilon f_{l_x}, \label{eq9a}\tag{9a} \\
			\lambda f_{l_y+1}-2\left[\cos{(2\pi \alpha_x l_y+\theta^{\prime})}+\cos{(2\pi \alpha_{y}l_y)}\right]f_{l_y} &=\varepsilon f_{l_y}, \label{eq9b}\tag{9b}
		\end{align}
		with $\theta=2\pi \alpha_y l$ and $\theta^{\prime}=2\pi \alpha_x l$. It is easy to see that the Eq. $\eqref{eq9a}$ and Eq. $\eqref{eq9b}$ show the similar structure. We study the topological properties of the 1D non-Hermitian chains described by Eq. $\eqref{eq9a}$ in the dual space, and one can obtain the same results by studying Eq. $\eqref{eq9b}$.
		Inspired by the definition of topological invariant of 1D non-Hermitian system \cite{HuitaoShen,Zongping},  we can introduce a winding number $w$ to characterize the topological feature of the above decoupled 1D equations, which is given as follows:
		\begin{equation}\label{eq10}
			w=\frac{1}{2\pi i} \int_0^{2\pi}d\phi \partial_{\phi}\ln{\det{\left[\mathcal{H}(\phi)-\varepsilon_B\right]}}=\frac{1}{2\pi}\int_{0}^{2\pi}d\phi \partial_{\phi} \Phi(\phi), \tag{10}
		\end{equation}
		where $\mathcal{H}$ is the corresponding Hamiltonian of Eq. $\eqref{eq9a}$ or Eq. $\eqref{eq9b}$ under periodic boundary conditions (PBCs), the phase $\phi$ is a magnetic flux penetrating through the center of a ring chain, $\varepsilon_B$ is a chosen base energy, and $\Phi(\phi)$ is the argument of $\det{\left[\mathcal{H}(\phi)-\varepsilon_B\right]}$. The winding number $w$ counts the times of the complex spectral trajectory encircling $\varepsilon_B$ with the rolling of the phase $\phi$ from $0$ to $2\pi$. Apparently, the winding number should vanish when the spectrum of the system is real. We introduce a magnetic flux $\phi$ penetrating through the non-Hermitian chain under PBCs such that the hopping amplitude between the boundaries is multiplied by $e^{i\phi}$ under a specific choice of gauge. As shown in Appendix A2, $\lambda_{c1}$ is the topological phase transition point, which is also the transition point of the extended regime and the intermediate regime with MEs. According to the definition of the winding number of the non-Hermitian system, the 2D QCs with $\mathcal{PT}$ symmetry emerge two distinct topological phases. One is the entirely extended eigenstates and $\mathcal{PT}$ symmetry unbroken regime with $w=0$ for $\lambda<\lambda_{c1}$, and the other is the emergence of localized eigenstates and $\mathcal{PT}$ symmetry broken regime with $w=1$ for $\lambda>\lambda_{c1}$. To check the validity of the analytical results, we perform numerical diagonalization calculation of the effective 1D case described by Eq. $\eqref{eq9a}$ with $\alpha_{x}=(\sqrt{5}-1)/2$, $\alpha_y=\sqrt{3}$, $L=1000$ and under PBCs. Figure \ref{Fig2}(a) shows the winding number $w$ of Eq. $\eqref{eq9a}$ as a function of $\lambda$ with $\varepsilon_B=0$, and the data are averaged $10000$ different $\theta$ realizations. The numerical results (circles) are expected from the analytical one (solid line). The inset of Fig. \ref{Fig2}(a) displays how $\Phi$ changes with $\phi$ from $-\pi$ to $\pi$ for different $\theta$ and $\lambda$ at either side of $\lambda_{c1}$. Figures \ref{Fig2}(b)-\ref{Fig2}(d) show the eigenvalues of Eq. $\eqref{eq9a}$ under PBCs with different $\lambda$, $L=1000$, and rolling $\theta$ from $0$ to $2\pi$. In $\mathcal{PT}$ symmetry unbroken regime, all the eigenvalues are real. When $\lambda\in (\lambda_{c1},\lambda_{c2})$, the complex eigenvalues emerge, and the eigenvalues are the mixture of real and complex values. With the increase of $\lambda$ above $\lambda_{c2}$, all the eigenvalues become complex. Figure \ref{Fig2}(d) shows the behavior of the maximum value of $|\mathrm{Im}(\varepsilon)|$ for $\theta$ varying from $-\pi$ to $\pi$ as a function of $\lambda$. We observe an abrupt increase of $\mathrm{max}\{|\mathrm{Im}(\varepsilon)|\}$ from zero to a finite value at $\lambda=\lambda_{c1}=1$, corresponding to the $\mathcal{PT}$ symmetry broken point predicted by the theoretical analysis. Note that the results correspond to the 2D non-Hermitian QCs in the thermodynamics limit, which displays a distinct $\mathcal{PT}$ symmetry phase transition. However, for the 2D non-Hermitian QCs with a finite size $L\times L$ described by Hamiltonian (\ref{eq1}), due to the finite-size effect, it always displays the finite imaginary parts of eigenvalues in the $\lambda<\lambda_{c1}$ region (see Appendix A3).
	
	\section{Excitation dynamics in an optical setup}
	In the experiments, one can implement the 2D $\mathcal{PT}$-symmetric quasiperiodic model in an optical setup by applying the coupled single-mode waveguides. Due to the nonvanishing overlap between the evanescent modes, light propagating along a waveguide can hop to its neighboring waveguides. The coupling constant $t$ can be controlled by modulating the spacing between the waveguides. In addition, the real and the imaginary parts of the $\mathcal{PT}$-symmetric propagation constant $\lambda_{mn}$ can be effectively controlled by modulating the real and imaginary parts of each waveguide's refraction index, respectively. Typically, by introducing a global gain factor, one can directly map the offset for the imaginary part of the refraction index on the potentials with the gain-loss profile. Hence, the paraxial wave propagation in a 2D modulated waveguide array of $L \times L$ coupled channels is described by an equation which is identical to our non-Hermitian $\mathcal{PT}$-symmetric model described by Eq. (\ref{eq1}), where the light propagation axis, $z$, plays the role of time, $i\partial_z \varphi_{m,n}=\hat{H}\varphi_{m,n}$ with $\varphi_{m,n}=\psi_{m,n}\exp{(i\varepsilon z)}$. The localization of such a model can be determined by detecting the excitation dynamics in such a photonic platform.
	
	\begin{figure}[tbp]
		\begin{center}
			\includegraphics[width=.5 \textwidth] {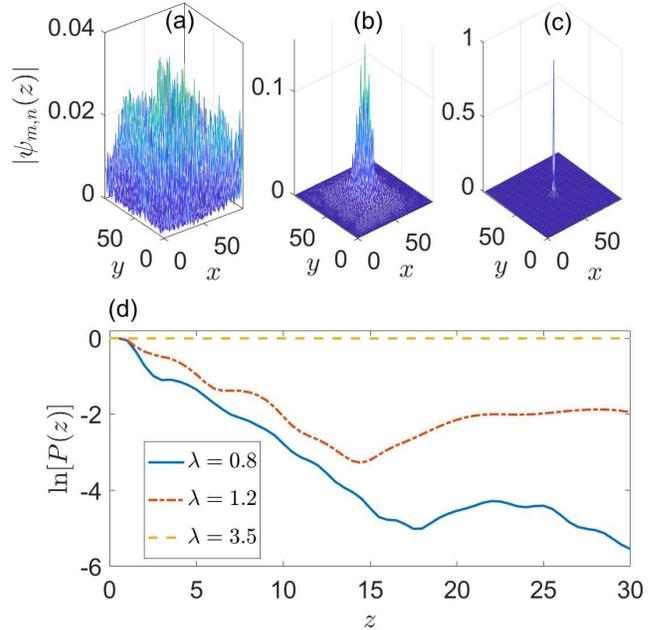}
		\end{center}
		\caption{(Color online) The outgoing profile distributions $|\psi_{m,n}(z)|$ at $z=30$ for (a) $\lambda=0.8$, (b) $\lambda=1.2$, and (c) $\lambda=3.5$, respectively. (d) The evolution of $P(z)$ for different $\lambda$ with the observed mesh's size in the center of the 2D system  $L^{\prime}\times L^{\prime}=5\time 5\times 5$.  Here, $\alpha_{x}=(\sqrt{5}-1)/2$, $\alpha_y=\sqrt{3}$ and $L\times L=99\times 99$.}\label{Fig3}
	\end{figure}
	
	To dynamically investigate the localization behaviors of the 2D non-Hermitian QCs, by injecting light into the center of the 2D waveguide system, one can obverse the evolution of light along the $z$-direction, {\it i.e.}, $|\psi(z)\rangle=1/\sqrt{\mathcal{N}}e^{-i\hat{H}z}|\psi(0)\rangle$, where $\mathcal{N}$ is the normalization coefficient of the wavefunction along the $z$-direction, and the initial state $|\psi(0)\rangle =\hat{c}^{\dagger}_{m_0,n_0}|0\rangle$ with $m_0$ and $n_0$ are the center of the lattice along the $x$- and $y$-directions, respectively. Figures. \ref{Fig3}(a)-\ref{Fig3}(c) show the profile of light in the waveguides at $z=30$ for different $\lambda$. The outgoing intensity displays a significant expansion for $\lambda=0.8$. When $\lambda=1.2$ in the regime with MEs, the intensity distribution's center remains a finite value, and the other part of the outgoing spreads is similar to that in the extended regime. For $\lambda=3.5$, deep in the localized regime, the outgoing intensity nearly freezes its position, shown a significant localized feature. We define $P(z)$, as the evolution of the probability of finding the excitation in the center mesh with the size $L^{\prime}\times L^{\prime} = 5\times 5$ \cite{ZhihaoXu2020,ZhihaoXuPRA,XDeng,Ketzmerick}, which is shown in Fig. \ref{Fig3}(d). When $\lambda$ is localized in the extended regime, $P(z)$ is a finite $z$ presents an exponential decay to a value $\propto (L^{\prime}/L)^2$. For a given $\lambda$ in the regime with MEs, $P(z)$ decays to a finite value, which depends on the portion of the localized states \cite{ZhihaoXu2020}. $P(z)$ is nearly frozen and approaches the unit for a localized regime. Different localization regimes show different features in the evolution of excitation dynamics.
	
	\section{Conclusions}
	
	We have proposed the 2D $\mathcal{PT}$ symmetric QC model with two incommensurate modulation frequencies along $x$- and $y$-directions, respectively, which displays exact MEs in the energy spectrum. In the thermodynamic limit, the localization transition point corresponding to the transition of the extended regime and the intermediate regime with MEs is topologically originating from the $\mathcal{PT}$ symmetry broken. A photonic waveguide setup can be used to realize the 2D non-Hermitian QCs, and the excitation dynamics can be used to detect the localization properties. Our study opens up a new way to investigate novel ME physics in 2D systems analytically.
	
	\begin{acknowledgements}
		Z.X. is supported by the NSFC (Grants No. 11604188 and No. 12047571), Beijing National Laboratory for Condensed Matter Physics, and STIP of Higher Education Institutions in Shanxi under Grant No. 2019L0097. X.X. is supported by Nankai Zhide Foundation. S.C. is supported by NSFC under Grant No. 11974413 and the Strategic Priority Research Program of the Chinese Academy of Sciences under Grant	No. XDB33000000. This work is also supported by NSF for	Shanxi Province Grant No. 1331KSC.
	\end{acknowledgements}

	\begin{appendix}	
		
		\renewcommand{\thesection}{Appendix}	
		
		\section{}

		\subsection{\label{sec:level1} Multifractal analysis}
		
		\begin{figure}[tbp]
			\begin{center}
				\includegraphics[width=.5 \textwidth] {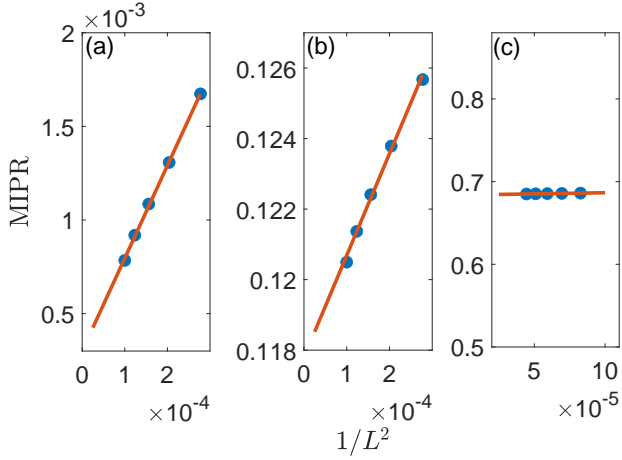}
			\end{center}
			\caption{(Color online) $\mathrm{MIPR}$ as a function of $1/L^2$ with (a) $\lambda=0.8$, (b) $\lambda=2.0$, and (c) $\lambda=3.5$, respectively. The solid line is a fitting line. Here, $\alpha_{x}=(\sqrt{5}-1)/2$, $\alpha_y=\sqrt{3}$, and under OBCs.}\label{SFig}
		\end{figure}
		
		To further strength the localization of the 2D non-Hermitian quasicrystals with $\alpha_x \ne \alpha_y$, we study the scaling behavior of eigenstates by performing a multifracal analysis. We define the mean inverse participation ratio (MIPR): $\mathrm{MIPR}=\sum_{j} \mathrm{IPR}_j /L^2$, where $\mathrm{IPR}_j$ represents the IPR of the $j$-th normalized eigenstate. Figure \ref{SFig} shows $\mathrm{MIPR}$ as a function of $1/L^2$ for different $\lambda$ with $\alpha_{x}=(\sqrt{5}-1)/2$, $\alpha_y=\sqrt{3}$ under OBCs. In the extended regime shown in Fig. \ref{SFig}(a) with $\lambda=0.8$, $\mathrm{MIPR} \propto L^{-2}$, and in the large size limit, the MIPR approaches to $0$. When $\lambda$ is localized in the intermediate regime, $\mathrm{MIPR}$ decays as $L^{-2}$ with the increase of $L$, and finial tends to a finite value about $0.1178$ with $\lambda=2.0$ [see Fig. \ref{SFig}(b)]. As shown in Fig. \ref{SFig}(c) with $\lambda=3.5$ in the localized regime, $\mathrm{MIPR}$ keeps a finite value about $0.6838$ with the increase of the size of the system.
		
		\subsection{\label{sec:level2} Winding number}
		To calculate the topological number of the 2D non-Hermitian quasicrystals described by Eq. (3) in the main text, we take a Fourier transformation
		\begin{equation}\label{seq1}
			f_{l_x,l_y}=\frac{1}{L}\sum_{m,n} e^{i 2\pi (\alpha_x l_x m+\alpha_y l_y n)} \psi_{m,n}. \tag{a1}
		\end{equation}
		In the thermodynamics limit, the dual equation can be written as
		\begin{equation}\label{seq2}
			\lambda f_{l_x+1,l_y+1}-2[\cos{2\pi\alpha_x l_x}+\cos{2\pi\alpha_y l_y}]f_{l_x,l_y}=\varepsilon f_{l_x,l_y}. \tag{a2}
		\end{equation}
		It is easy to find that the 2D dual equation can be reduced to 1D chains along the diagonal direction, and these 1D chains are decoupled. Hence, for a infinite 2D system, it can be equivalent to a serial of 1D chains labeled by different $l=l_x-l_y$. For a given $l$, the eigenvalue equation of the 1D chain is given as follows:
		\begin{equation}\label{seq3}
			\lambda f_{l_x+1}-2[\cos{(2\pi\alpha_x l_x)}+\cos{(2\pi\alpha_y l_x-\theta)}]f_{l_x}=\varepsilon f_{l_x}, \tag{a3}
		\end{equation}
		with $\theta=2\pi \alpha_y l$. The winding number of the 1D non-Hermitian chain is defined as
		\begin{equation}\label{seq4}
			w=\frac{1}{2\pi i}\int_{0}^{2\pi} d\phi \partial_{\phi} \ln{\det{[\mathcal{H}(\phi)-\varepsilon_B]}}, \tag{a4}
		\end{equation}
		where $\mathcal{H}$ is the Hamiltonian of the 1D non-Hermitian chain under the periodic boundary condition, the phase $\phi$ is considered as a magnetic flux through a non-Hermitian ring, and the base frequency $\varepsilon_B=0$ is chosen. The matrix form of the 1D non-Hermitian chain with a magnetic flux is written as
		\begin{equation}\label{seq5}
			\mathcal{H}(\phi)=\begin{pmatrix} -V_1 & \lambda & 0 & 0 & \cdots & 0 & 0 \\
				0 & -V_2 & \lambda & 0 & \cdots & 0 & 0\\
				0 & 0 & -V_3 & \lambda & \cdots & 0 & 0\\
				\vdots & \vdots  & \vdots & \vdots & \vdots & \vdots & \vdots\\
				0 & 0 & 0 & 0 & \cdots & -V_{L-1} & \lambda\\
				\lambda e^{i\phi} & 0 & 0 & 0 & \cdots & 0 & -V_L \\
			\end{pmatrix}, \tag{a5}
		\end{equation}
		with $V_{l_x}=2[\cos{(2\pi\alpha_x l_x)}+\cos{(2\pi\alpha_y l_x-\theta)}]$. Hence, the determination of $\mathcal{H}(\phi)$
		\begin{equation}\label{seq6}
			D(\phi)=\det{[\mathcal{H}(\phi)]}=(-1)^L\left(\prod_{l_x=1}^{L}V_{l_x}-\lambda^L e^{i\phi}\right). \tag{a6}
		\end{equation}
		Since the winding number of the non-Hermitian case describes the complex spectral trajectory encircles the base point $\omega_B=0$ when the flux varies from $0$ to $2\pi$, we can rewrite the winding number as follows:
		\begin{equation}\label{seq7}
			w=\frac{1}{2}\sum_{j} \mathrm{sgn}\left\{\mathrm{Re}[D(\phi_j)]\right\}\cdot \mathrm{sgn} \left\{\frac{d \mathrm{Im}[D(\phi_j)]}{d\phi}\right\}, \tag{a7}
		\end{equation}
		where $\mathrm{Re}[D(\phi)]=(-1)^L\left(\prod_{l_x=1}^{L}V_{l_x}-\lambda^L\cos{\phi}\right)$, $\mathrm{Im}[D(\phi)]=(-1)^{L+1}\lambda^L\sin{\phi}$, and $\mathrm{sgn}(\cdots)$ denotes the sign operator. $\phi_j$ is the $j$th solution of $\mathrm{Im}[D(\phi)]=0$ with $\phi_1=0$ and $\phi_2=\pi$, respectively. We have
		\begin{align}\label{seq8}
			w &=\frac{1}{2}\left[\mathrm{sgn}\left( \prod_{l_x=1}^{L}V_{l_x}+\lambda^L\right)- \mathrm{sgn}\left( \prod_{l_x=1}^{L}V_{l_x}-\lambda^L\right) \right] \notag \\
			&= S\left(\lambda^L-\left|\prod_{l_x=1}^{L}V_{l_x}\right|\right), \tag{a8}
		\end{align}
		with the step function
		\begin{equation}\label{seq9}
			S(x) = \begin{cases} 0, \quad x<0 \\
				1/2,\quad x=0 \\
				1,\quad x>0.
			\end{cases} \tag{a9}
		\end{equation}
		The topological phase transition point is determined by
		\begin{equation}\label{seq10}
			\lambda^L=\left|\prod_{l_x=1}^{L} V_{l_x}\right|. \tag{a10}
		\end{equation}
		In the large $L$ limit, we have
		\begin{align}\label{seq11}
			\ln{\lambda} = &\lim_{L\to\infty} \frac{1}{L} \sum_{l_x=1}^{L}\ln{|V_{l_x}|} \notag \\
			= & \int_{0}^{1} \ln{\left|2[\cos{(2\pi\alpha_x Lx)}+\cos{(2\pi\alpha_y Lx-\theta)}]\right|} dx \notag \\
			= & 2\ln{2}+\int_{0}^{1} \ln{\left|\cos{\frac{2\pi(\alpha_x+\alpha_y)Lx-\theta}{2}}\right|} dx \notag  \\
			& + \int_{0}^{1} \ln{\left|\cos{\frac{2\pi(\alpha_x-\alpha_y)Lx+\theta}{2}}\right|} dx \notag  \\
			= & 2\ln{2}+\frac{1}{\pi(\alpha_x+\alpha_y)L}\mathcal{L}\left[\pi(\alpha_x+\alpha_y)L\right] \notag \\
			& +\frac{1}{\pi(\alpha_x -\alpha_y)L}\mathcal{L}\left[\pi(\alpha_x-\alpha_y)L\right] \notag \\
			\approx & 0, \tag{a11}
		\end{align}
		where
		\begin{align}\label{seq12}
			\mathcal{L}(x) &=\int_0^{x} \ln{\cos{(x^{\prime})}}dx^{\prime} \notag \\
			&=-x\ln{2}+\frac{1}{2}\sum_{k=1}^{\infty}(-1)^{k-1}\frac{\sin(2kx)}{k^2}. \tag{a12}
		\end{align}
		It means that in thermodynamic limit $L\to \infty$, the topological phase transition point at $\lambda_{c1}=e^{0}=1$. 	
		
		\subsection{\label{sec:level3} Energy spectrum of the 2D non-Hermitian quasicrystals with the finite size}
		\begin{figure}[tbp]
			\begin{center}
				\includegraphics[width=.5 \textwidth] {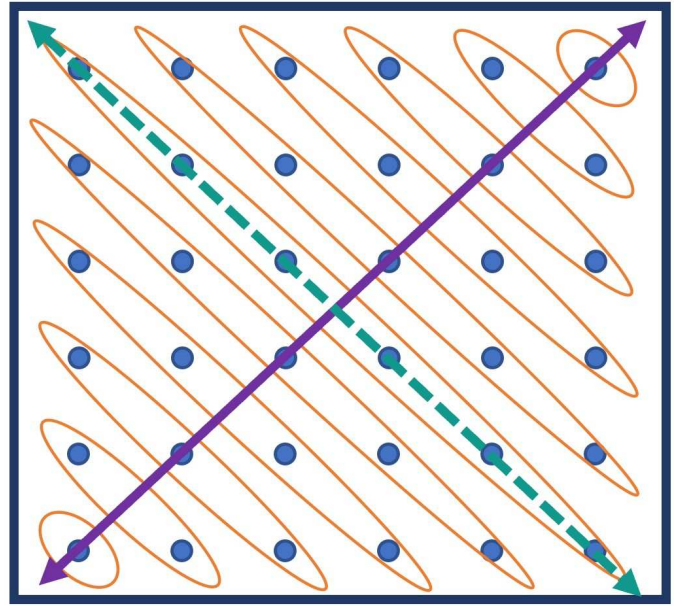}
			\end{center}
			\caption{(Color online) Schematic of a finite 2D quasicrystal in the dual space for our model. The arrows with the dashed green lines denote the diagonal directions and the arrows with the solid purple lines represent the anti-diagonal directions.}\label{SFigS}
		\end{figure}
		
		\begin{figure}[tbp]
			\begin{center}
				\includegraphics[width=.5 \textwidth] {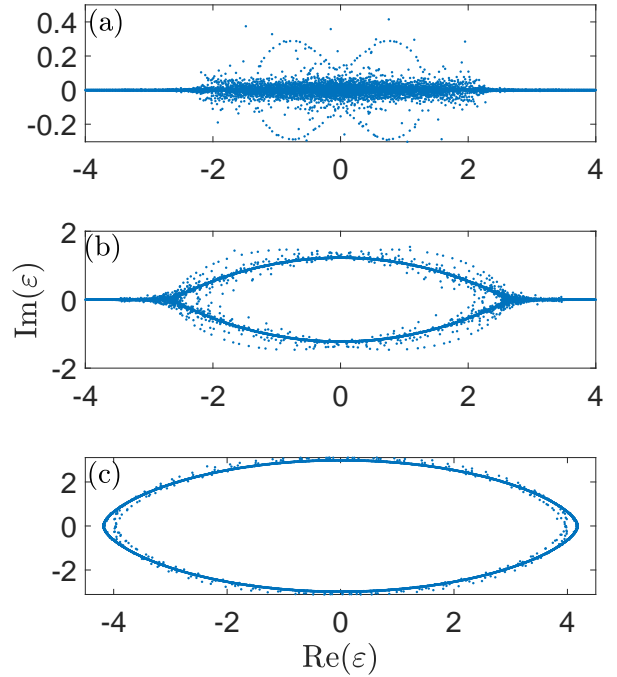}
			\end{center}
			\caption{(Color online) Eigenvalues of the 2D non-Hermitian quasicrystal with (a) $\lambda=0.8$, (b) $\lambda=2.0$, and (c) $\lambda=3.5$, respectively. Here, $\alpha_{x}=(\sqrt{5}-1)/2$, $\alpha_y=\sqrt{3}$, $L\times L=80\times 80$, and under OBCs.}\label{SFig1}
		\end{figure}
		
		\begin{figure}[tbp]
			\begin{center}
				\includegraphics[width=.5 \textwidth] {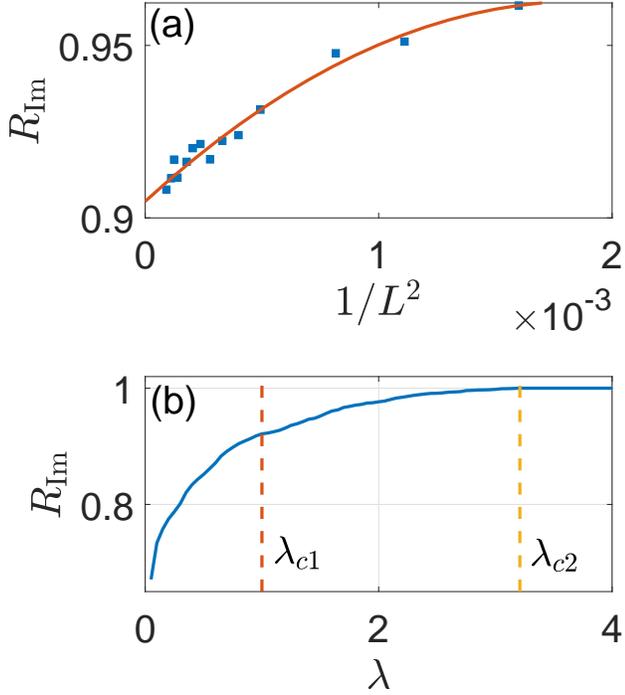}
			\end{center}
			\caption{(Color online) (a) $R_{\mathrm{Im}}$ as a function of $1/L^2$ with $\lambda=0.8$. The solid line is a fitting line. (b) $R_{\mathrm{Im}}$ vs $\lambda$ with $L\times L=80\times 80$.  Here, $\alpha_{x}=(\sqrt{5}-1)/2$, $\alpha_y=\sqrt{3}$, and under OBCs.}\label{SFig2}
		\end{figure}
		
		In the main text, we study a real-complex transition of the energy spectrum for a 2D non-Hermitian quasicrystal in the large $L$ limit. For our model with an infinite size, the system in the dual space can be reduced to a serial of decoupled chains along the diagonal direction labeled by $l=l_x-l_y$. Our results imply that the real-complex transition point corresponds to the topological phase transition point $\lambda_{c1}$ in the thermodynamic limit. However, for an arbitrary finite system $L\times L$, a heuristic argument is that in dual space the length of the 1D chains shrinks along the anti-diagonal direction from the center to the corners, and finally, the 1D chain turns to a single lattice at the corners (see Fig. \ref{SFigS}). It implies the finite-size effect is remarkable in our system. Unlike in the infinite case that all the eigenvalues are real in $\lambda<\lambda_{c1}=1$, a finite number of complex energies is observed, which is shown in Fig. \ref{SFig1}(a) with $L\times L=80\times 80$ and $\lambda=0.8$ under OBCs. Due to the finite-size effect, parts of eigenvalues deviate from that in the thermodynamic limit, as shown in Figs. \ref{SFig1}(b) and \ref{SFig1}(c). We define $R_{\mathrm{Im}}=N_{\mathrm{Im}}/L^2$, where $N_{\mathrm{Im}}$ is the number of eigenvalues with the modulus of the imaginary parts above the cutoff $10^{-6}$. Figures \ref{SFig2}(a) shows $R_{\mathrm{Im}}$ as a function of $1/L^2$ with $\lambda=0.8$ under OBCs. The results imply that in the current geometric configuration, the number of the complex eigenvalues of the finite 2D non-Hermitian quasicrystal keeps finite. We also calculate the behavior of $R_{\mathrm{Im}}$ versus $\lambda$. For $\lambda<\lambda_{c2}$, $R_{\mathrm{Im}}$ displays a monotonous increase, and when $\lambda\ge \lambda_{c2}$, $R_{\mathrm{Im}}=1$ which is shown in Fig. \ref{SFig2}(b).

	\end{appendix}


\begin{thebibliography}{99}
		\bibitem{Anderson} P. W. Anderson, Phys. Rev. \textbf{109}, 1492 (1958).
		\bibitem{Cutler} M. Cutler and N. F. Mott, Phys. Rev. \textbf{181}, 1336 (1969).
		\bibitem{Ramakrishnan} P. A. Lee and T. V. Ramakrishnan, Rev. Mod. Phys. \textbf{57}, 287 (1985).
		\bibitem{Dalichaouch} R. Dalichaouch, J. P. Armstrong, S. Schultz, P. M. Platzman, and S. L. McCall, Nature (London)	\textbf{354}, 52 (1991).
		\bibitem{Wiersma} D. S. Wiersma, P. Bartonlini, A. Lagendijk, and R. Righini, Nature (London) \textbf{390}, 671 (1997).
		\bibitem{Storzer} M. St\"{o}rzer, P. Gross, C. M. Aegerter, and G. Maret, Phys. Rev. Lett. \textbf{96}, 063904	(2006).
		\bibitem{Topolancik} J. Topolancik, B. Ilic, and F. Vollmer, Phys. Rev. Lett. \textbf{99}, 253901 (2007).
		\bibitem{Schwartz} T. Schwartz, G. Bartal, S. Fishman, and M. Segev, Nature (London) \textbf{446}, 52 (2007).
		\bibitem{Strybulevych} H. Hu, A. Strybulevych, J. H. Page, S. E. Skipetrov, and B. A. van Tiggelen, Nat. Phys. \textbf{4}, 946 (2008).
		\bibitem{Chabe} J. Chab\'{e}, G. Lemari\'{e}, B. Gr\'{e}maud, D. Delande, P. Szriftgiser, and J. C. Garreau, Phys. Rev. Lett. \textbf{101}, 255702 (2008).
		\bibitem{Sperling} T. Sperling, W. B\"{u}hrer, C. M. Aegerter, and G. Maret, Nat. Photonics \textbf{7}, 48 (2012).
		\bibitem{Lopez} M. Lopez, J.-F. Szriftgiser, J. C. Garreau, and D. Delande, Phys. Rev. Lett. \textbf{108}, 095701 (2012).
		\bibitem{Manai} I. Manai, J.-F. Cl\'{e}ment, R. Chicireanu, C. Hainaut, J. C. Garreau, P. Szriftgiser, and D. Delande, Phys. Rev. Lett. \textbf{115}, 240603 (2015).
		\bibitem{Tianping} T. Ying, Y. Gu, X. Chen, X. Wang, S. Jin, L. Zhao, W. Zhang, and X. Chen, Sci. Adv. \textbf{2}, e1501283 (2016).
		\bibitem{Hutchinson} D. H. White, T. A. Haase, D. J. Brown, M. D. Hoogerland, M. S. Najafabadi, J. L. Helm, C. Gies, D. Schumayer, and
		D. A. W. Hutchinson, Nat. Commun. \textbf{11}, 4941 (2020).
		\bibitem{Abrahams} E. Abrahams, P. W. Anderson, D. C. Licciardello, and T. V. Ramakrishnan, Phys. Rev. Lett. \textbf{42}, 673 (1979).
		\bibitem{Evers} F. Evers and A. D. Mirlin, Rev. Mod. Phys. \textbf{80}, 1355 (2008).
		\bibitem{Whitney} R. Whitney, Most Efficient, Phys. Rev. Lett. \textbf{112}, 130601 (2014).
		\bibitem{Chiaracane} C. Chiaracane, M. T. Mitchison, A. Purkayastha, G. Haack, and J. Goold, Phys. Rev. Research \textbf{2}, 013093 (2020).
		\bibitem{AA} S. Aubry and G. Andr\'{e}, Ann. Isr. Phys. Soc. \textbf{3}, 133 (1980).
		\bibitem{Sarma1988} S. Das Sarma, S. He, and X. C. Xie, Phys. Rev. Lett. \textbf{61}, 2144 (1988).
		\bibitem{Boers} D. J. Boers, B. Goedeke, D. Hinrichs, and M. Holthaus, Phys. Rev. A \textbf{75}, 063404 (2007).
		\bibitem{Biddle2009} J. Biddle, B. Wang, D. J. Priour Jr, and S. Das Sarma, Phys. Rev. A \textbf{80}, 021603(R) (2009).
		\bibitem{Biddle2010} J. Biddle and S. Das Sarma, Phys. Rev. Lett. \textbf{104}, 070601 (2010).
		\bibitem{Lellouch} S. Lellouch and L. Sanchez-Palencia, Phys. Rev. A \textbf{90}, 061602(R) (2014).
		\bibitem{XLi2017} X. Li, X. Li, and S. Das Sarma, Phys. Rev. B \textbf{96}, 085119 (2017).
		\bibitem{HYao2019}  H. Yao, H. Khoudli, L. Bresque, and L. Sanchez-Palencia, Phys. Rev. Lett. \textbf{123}, 070405 (2019).
		\bibitem{Ganeshan} S. Ganeshan, J. H. Pixley, and S. Das Sarma, Phys. Rev. Lett. \textbf{114}, 146601 (2015).
		\bibitem{ZhihaoXu2020}  Z. Xu, H. Huangfu, Y. Zhang, and S. Chen, New J. Phys. \textbf{22}, 013036 (2020).
		\bibitem{YLiu2020} Y. Liu, X.-P. Jiang, J. Cao, and S. Chen, Phys. Rev. B \textbf{101}, 174205 (2020).
		\bibitem{LiuPRB2021} Y. Liu, Q. Zhou, and S. Chen, Phys. Rev. B \textbf{104}, 024201 (2021). 
		\bibitem{QBZeng} Q.-B. Zeng and Y. Xu, Phys. Rev. Research \textbf{2}, 033052 (2020).
		\bibitem{Longhi2021} S. Longhi, arXiv:2106.00230.
		\bibitem{XuXia} X. Xia, K. Huang, S. Wang, X. Li, arXiv:2105.12640.
		\bibitem{YanxiaoLiu2021} Y. Liu, Y. Wang, Z. Zheng, and S. Chen, Phys. Rev. B \textbf{103}, 134208 (2021).
		\bibitem{LingZhiTang} L.-Z. Tang, G.-Q. Zhang, L.-F. Zhang, D.-W. Zhang, Phys. Rev. A 103, 033325 (2021)
		\bibitem{LiuTong} T. Liu, X. Xia, S. Longhi, and L. Sanchez-Palencia, arXiv:2105.04591.
		\bibitem{LonghiPRB2021} S. Longhi, Phys. Rev. B \textbf{103}, 144202 (2021).
		\bibitem{LonghiPRB20211} S. Longhi, Phys. Rev. B \textbf{103}, 054203 (2021).
		\bibitem{TongLiu2020} T. Liu, H. Guo, Y. Pu, and S. Longhi, Phys. Rev. B \textbf{102}, 024205 (2020).
		\bibitem{LonghiPRB2019} S. Longhi, Phys. Rev. B \textbf{100}, 125157 (2019).
		\bibitem{LonghiPRL2019} S. Longhi, Phys. Rev. Lett. \textbf{122}, 237601 (2019).
		\bibitem{HJiang} H. Jiang, L.-J. Lang, C. Yang, S.-L. Zhu, and S. Chen, Phys. Rev. B \textbf{100}, 054301 (2019).
		\bibitem{XDeng} X. Deng, S. Ray, S. Sinha, G. V. Shlyapnikov, and L.
		Santos, Phys. Rev. Lett. \textbf{123}, 025301 (2019).
		\bibitem{Yucheng} Y. Wang, X. Xia, L. Zhang, H. Yao, S. Chen, J. You, Q. Zhou, and X.-J. Liu, Phys. Rev. Lett. \textbf{125}, 196604 (2020).
		\bibitem{ZhihaoXuPRA} Z. Xu and S. Chen, Phys. Rev. A \textbf{103}, 043325 (2021).
		\bibitem{YuCheng2021} Y. Wang, X. Xia, Y. Wang, Z. Zheng, and X.-J. Liu, Phys. Rev. B \textbf{103}, 174205(2021).		
		\bibitem{Bender} C. M. Bender and S. Boettcher, Phys. Rev. Lett. \textbf{80}, 5243 (1998).
		\bibitem{Bender1} C. M. Bender, Rep. Prog. Phys. \textbf{70}, 947 (2007).
		\bibitem{Hatano1} N. Hatano and D. R. Nelson, Phys. Rev. Lett. \textbf{77}, 570 (1996).
		\bibitem{Hatano2} N. Hatano and D. R. Nelson, Phys. Rev. B \textbf{56}, 8651 (1997).
		\bibitem{Hatano3} N. Hatano and D. R. Nelson, Phys. Rev. B \textbf{58}, 8384 (1998).
		\bibitem{SLongi1} S. Longhi, Ann. Phys. (Berlin) \textbf{530}, 1800023 (2018).
		\bibitem{HuitaoShen} H. Shen, B. Zhen, and L. Fu, Phys. Rev. Lett. \textbf{120}, 146402 (2018).
		\bibitem{Shunyu1} S. Yao and Z. Wang, Phys. Rev. Lett. \textbf{121}, 086803 (2018).
		\bibitem{Shunyu2} S. Yao, F. Song, and Z. Wang, Phys. Rev. Lett. \textbf{121}, 136802 (2018).
		\bibitem{Zongping} Z. Gong, Y. Ashida, K. Kawabata, K. Takasan, S. Higashikawa, and M. Ueda, Phys. Rev. X \textbf{8}, 031079 (2018).
		\bibitem{Lee1} T. E. Lee, Phys. Rev. Lett. \textbf{116}, 133903 (2016).
		\bibitem{Lee2} T. E. Lee and C.-K. Chan, Phys. Rev. X \textbf{4}, 041001 (2014).
		\bibitem{Harari} G. Harari, M. A. Bandres, Y. Lumer, M. C. Rechtsman, Y. D. Chong, M. Khajavikhan, D. N. Christodoulides, and M. Segev, Science \textbf{359}, eaar4003 (2018).
		\bibitem{Parto} M. Parto, S. Wittek, H. Hodaei, G. Harari, M. A. Bandres, J. Ren, M. C. Rechtsman, M. Segev, D. N. Christodoulides, and M. Khajavikhan, Phys. Rev. Lett. \textbf{120}, 113901 (2018).
		\bibitem{HengyunZhou} H. Zhou, C. Peng, Y. Yoon, C. W. Hsu, K. A. Nelson, L. Fu, J. D. Joannopoulos, M. Soljacic, and B. Zhen, Science \textbf{359}, 1009 (2018).
		\bibitem{Ruter} C. E. R\"{u}ter, K. G. Makris, R. El-Ganainy, D. N. Christodoulides, M. Segev, and D. Kip, Nat. Phys. \textbf{6}, 192 (2010).
		\bibitem{LiangFeng} L. Feng, M. Ayache, J. Huang, Y.-L. Xu, M.-H. Lu, Y.-F. Chen, Y. Fainman, and A. Scherer, Science \textbf{333}, 729 (2011).
		\bibitem{Regensburger} A. Regensburger, C. Bersch, M.-A. Miri, G. Onishchukov, D. N. Christodoulides, and U. Peschel, Nature (London) \textbf{488}, 167 (2012).
		\bibitem{Joglekar} J. Li, A. K. Harter, J. Liu, L. de Melo, Y. N. Joglekar, and L. Luo, Nat. Commun. \textbf{10}, 855 (2019).
		\bibitem{XiangZhan} X. Zhan, L. Xiao, Z. Bian, K. Wang, X. Qiu, B. C. Sanders, W. Yi, and P. Xue, Phys. Rev. Lett. \textbf{119}, 130501 (2017).
		\bibitem{LeiXiao} L. Xiao, X. Zhan, Z. H. Bian, K. K. Wang, X. Zhang, X. P. Wang, J. Li, K. Mochizuki, D. Kim, N. Kawakami, W. Yi, H. Obuse, B. C. Sanders, and P. Xue, Nat. Phys. \textbf{13}, 1117 (2017).
		\bibitem{Zeuner} J. M. Zeuner, M. C. Rechtsman, Y. Plotnik, Y. Lumer, S. Nolte, M. S. Rudner, M. Segev, and A. Szameit, Phys. Rev. Lett. \textbf{115}, 040402 (2015).
		\bibitem{YongXu} Y. Xu, S.-T. Wang, and L.-M. Duan, Phys. Rev. Lett. \textbf{118}, 045701 (2017).
		\bibitem{Okuma} N. Okuma and M. Sato, Phys. Rev. Lett. \textbf{123}, 097701 (2019).
		\bibitem{LinhuLi} C. H. Lee, L. Li, and J. Gong, Phys. Rev. Lett. \textbf{123}, 016805 (2019).
		\bibitem{Kunst} F. K. Kunst, E. Edvardsson, J. C. Budich, and E. J. Bergholtz, Phys. Rev. Lett. \textbf{121}, 026808 (2018).
		\bibitem{Takata} K. Takata and M. Notomi, Phys. Rev. Lett. \textbf{121}, 213902 (2018).
		\bibitem{MPan} M. Pan, H. Zhao, P. Miao, S. Longhi, and L. Feng, Nat. Commun. \textbf{9}, 1308 (2018).
		\bibitem{Nakagawa} M. Nakagawa, N. Kawakami, and M. Ueda, Phys. Rev. Lett. \textbf{121}, 203001 (2018).
		\bibitem{Hamazaki} R. Hamazaki, K. Kawabata, and M. Ueda, Phys. Rev. Lett. \textbf{123}, 090603 (2019).
		\bibitem{Yamamoto} K. Yamamoto, M. Nakagawa, K. Adachi, K. Takasan, M. Ueda, and N. Kawakami, Phys. Rev. Lett. \textbf{123}, 123601 (2019).
		\bibitem{Ashida} Y. Ashida, S. Furukawa, and M. Ueda, Nat. Commun. \textbf{8}, 15791 (2017).
		\bibitem{Kawabata} K. Kawabata, S. Higashikawa, Z. Gong, Y. Ashida, and M. Ueda, Nat. Commun. \textbf{10}, 297 (2019).
		\bibitem{Kawabata2} K. Kawabata, K. Shiozaki, M. Ueda, and M. Sato, Phys. Rev. X \textbf{9}, 041015 (2019).
		\bibitem{Zirnstein} H.-G. Zirnstein, G. Refae, and B. Rosenow, Phys. Rev. Lett. \textbf{126}, 216407 (2021).
		\bibitem{Okuma1} N. Okuma and M. Sato, Phys. Rev. Lett. \textbf{126}, 176601 (2021).
		\bibitem{Longwen1} L. Zhou and W. Han, arXiv:2105.03302.
		\bibitem{Longwen2} L. Zhou, Phys. Rev. Res. \textbf{3}, 033184 (2021).
		\bibitem{BHuang} B. Huang and W. Vincent Liu, Phys. Rev. B \textbf{100}, 144202 (2019).
		\bibitem{Pupillo} A. Gei$\beta$ler and G. Pupillo, Phys. Rev. Res. \textbf{2}, 042037(R) (2020).
		\bibitem{Schneider} A. Szab\'{o} and U. Schneider, Phys. Rev. B \textbf{101}, 014205 (2020).
		\bibitem{Asatryan} A. A. Asatryan, L. C. Botten, M. A. Byrne, R. C. McPhedran, and C. M. de Sterke, Phys. Rev. E \textbf{75}, 015601(R) (2007).
		\bibitem{Sarnak} P. Sarnak, Commun. Math. Phys. \textbf{84}, 377 (1982).
		\bibitem{Ketzmerick} R. Ketzmerick, K. Kruse, S. Kraut, and T. Geisel, Phys. Rev. Lett. \textbf{79}, 1959 (1997).		
		
	\end{thebibliography}
\end{document}